\documentclass{article}

\PassOptionsToPackage{numbers, compress, sort}{natbib}

\usepackage[nonatbib,preprint]{neurips_2019}

\usepackage[utf8]{inputenc} 
\usepackage[T1]{fontenc}    
\usepackage{hyperref}       
\usepackage{url}            
\usepackage{booktabs}       
\usepackage{amsfonts}       
\usepackage{nicefrac}       
\usepackage{microtype}      
\usepackage{cite}
\usepackage{amsmath,amssymb,amsfonts}
\usepackage{algorithmic}
\usepackage{graphicx}
\usepackage{textcomp}
\usepackage{subfigure}
\usepackage[linesnumbered,ruled,noend]{algorithm2e}
\title{HetuMoE: An Efficient Trillion-scale Mixture-of-Expert Distributed Training System}

\author{%
  Xiaonan Nie, Pinxue Zhao, Xupeng Miao, Tong Zhao, Bin Cui\\
School of Computer Science,\\
Peking University, Beijing, China \\
  \texttt{\{xiaonan.nie, 1800017766, xupeng.miao, zhaotong, bin.cui\}@pku.edu.cn} \\
}

\begin{document}

\maketitle

\begin{abstract}
As giant dense models advance quality but require large amounts of GPU budgets for training, the sparsely gated Mixture-of-Experts (MoE), a kind of conditional computation architecture, is proposed to scale models while keeping their computation constant. 
Specifically, the input tokens are routed by the gate network and only activates part of the expert network.
Existing MoE training systems only support part of mainstream MoE models (e.g. Top k) training under expensive high-bandwidth GPU clusters.
In this paper, we present HetuMoE, a high-performance large-scale sparse MoE training system built on Hetu. HetuMoE provides multiple gating strategies and efficient GPU kernel implementations. To further improve the training efficiency on commodity GPU clusters (e.g, with only 1 NiC), we introduce the hierarchical AllToAll communication that combines hierarchical networks and aggregating messages.
Compared with existing state-of-the-art MoE systems, HetuMoE obtains at least $15\%$ speedup. Specifically, HetuMoE outperforms DeepSpeed-MoE up to $8.1\times$ under the switch gate with a batch size of 32. Our code is available at: \url{https://github.com/PKU-DAIR/Hetu}.
\end{abstract}

\section{Introduction}
Models tend to perform better with increasing data size, parameter size in many fields, such as natural language processing (NLP) and computer vision (CV)~\cite{scalinglaws, bert, gpt2, t5, vit}.
However, these large models always require huge amounts of GPU resources and take weeks or even months to train. For example, it takes approximately 288 years with a single V100 NVIDIA GPU to train the GPT-3 model with 175 billion parameters~\cite{gpt3,megatron-lm}.
The sparsely gated Mixture of Experts (MoE), as a kind of conditional computation architecture, has been proved to be an effective way to expand model size without significantly increasing the computation~\cite{shazeer2017outrageously,fedus2021switch,GShard, vmoe,du2021glam}.
Specifically, an MoE model contains a gating network and a pool of experts. During training, a token will be assigned to a small number of experts for computation. This kind of sparsely-activated nature enables MoE to significantly expand the model size while keeping the amount of computation almost constant.
The sparsely-activated training paradigm necessitates new systems' support.
However, existing MoE training systems, including DeepSpeed-MoE~\cite{rajbhandari2022deepspeed}, Tutel~\cite{tutel}, and FastMoE~\cite{he2021fastmoe}, are still facing some limitations in both usability and efficiency. First, they only support parts of mainstream MoE models and gate networks (e.g. Top k). Second, most of them assume ``hyperclusters'' that have expensive, high-speed interconnects such as NVLink or Infiniband. They are suffering from a severe communication bottleneck for common clusters. The above shortcomings have restricted their usage and limited the exploration of MoE models in real applications.


In this paper, we first briefly introduce the common training process of MoE models and then propose HetuMoE, a high-performance distributed MoE training system built on Hetu, which supports a variety of mainstream gating strategies and achieves state-of-the-art training speed compared to existing baseline systems. HetuMoE adopts several customized gating kernels' implementation and utilizes the hierarchical All-To-All communication optimization.
Experiments show that HetuMoE outperforms existing MoE systems at least a $15\%$ speedup under different settings. Meanwhile, HetuMoE outperforms DeepSpeed-MoE up to $8.1\times$ under the switch gate with the  batch size of 32.

\section{MoE Training Process}
An MoE model consists of a gating network and a pool of sub-networks (e.g. Feed-Forward Networks in Transformer), which are called experts.
We formulate the common training process of MoE models in Algorithm~\ref{alg:moe_training}.

\begin{enumerate}
    \item [(1)] The input tokens are first processed by the gate network to know the target experts.
    \item [(2)] The layout transform operation is conducted on each device to put tokens with the same target experts in a continuous memory buffer for the incoming communication. (Line 4)
    \item [(3)] An AllToAll communication operation is incurred to dispatch tokens to their corresponding experts. (Line 6)
    \item [(4)] Then each expert process its tokens respectively. (Line 8-11)
    \item [(5)] The processed tokens are dispatched back to their GPUs. (Line 13)
    \item [(6)] The reverse layout transform operation is conducted to put tokens back to their original position in the training batch for future computation. (Line 15)
\end{enumerate}


\begin{algorithm}[t]
	\SetAlgoLined
	\SetKwProg{Fn}{Function}{}{end}
	\KwData{$x_{S}$: a group of tokens of size $S$, \\ 
	\quad \quad $Gate$: gate network \\
	\quad \quad $E$: expert network \\}
	\KwResult{$y_{S}$: tokens after precessing}
	\SetInd{0.61em}{0.61em}
	//Step 1: Get the target experts and corresponding weights. \;
	$\mathcal{W}_{(S, \ E)}, id_{S} = Gate(x_{S})$ \;
	//Step 2: Layout Transform on Input Data \;
	$x_{S}' = Layout\_Transform(x_{S}, id_{S})$ \;
	//Step 3: AllToAll Communication \;
	$x_{S}' = AllToAll(x_{S}')$ \;
	//Step 4: Expert Processing \;
	\For{$i \leftarrow 1\ to\ S$}{
	    $y_{s} = 0$ \;
	    \For{$idx \in id_{i}$}{
            $y_{s} = y_{s} + w_{(i, \ idx)} * e_{idx}(x_{i})$ \;
        }
    }
	//Step 5: AllToAll Communication \;
	$y_{S} = AllToAll(y_{S})$ \;
	//Step 6: Reverse Layout Transform on Output Data \;
	$y_{S} = Reverse\_Layout\_Transform(y_{S}, id_{S})$ \;
\caption{General MoE Training Process}
\label{alg:moe_training}
\end{algorithm}

\begin{figure}[htp]
\centering 
\includegraphics[width = 4.5cm]{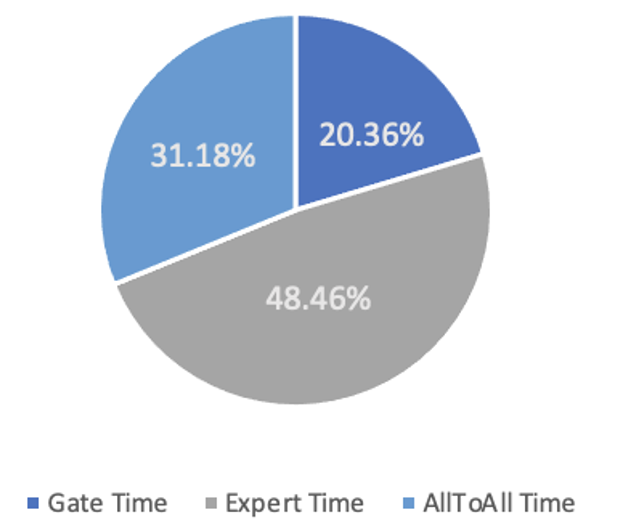}
\caption{Time consumption of MoE layer of DeepSpeed-MoE under 8 A100 GPUs in a single node}
\label{fig:time_percentage}
\end{figure}

We then evaluate DeepSpeed-MoE and profile its time costs of MoE layer under 8 A100 GPUs in a single node and the result is shown in Figure~\ref{fig:time_percentage}.
As we can see, the computation of gate network, including layout transform and its reverse operation, and AllToAll communication account for more than 50\% of training time totally. When scaling training into distributed scenarios across multiple nodes, AllToAll communication overheads would deteriorate the entire training process, which takes around 99\% of training time under a 100 Gbps network.

\begin{figure*}[t]
\centering 
\includegraphics[width=\columnwidth]{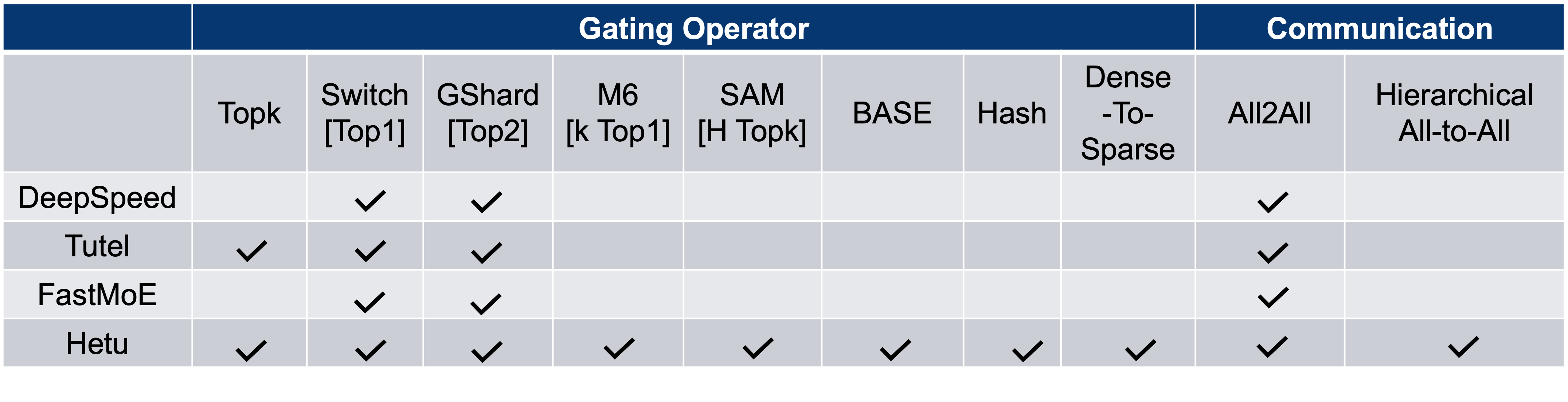}
\caption{HetuMoE compared with other MoE systems}
\label{fig:sys_compare}
\end{figure*}

\section{HetuMoE}
In order to solve above problems, we develop HetuMoE based on Hetu~\footnote{https://github.com/PKU-DAIR/Hetu}, a high-performance distributed deep learning system. HetuMoE supports various gating strategies proposed by recent MoE approaches, such as Switch~\cite{fedus2021switch}, GShard~\cite{GShard}, 
M6~\cite{M6T}, 
BASE Layer~\cite{BaseLayer} , 
Hash Layer~\cite{HashLayer}, 
SAM~\cite{SAM},
and Dense-to-Sparse~\cite{dense_to_sparse}. In addition, we implement hierarchical AllToAll~\cite{ha2a}, which greatly improves network bandwidth utilization in the case of multi-node distributed training under commodity network conditions, compared with vanilla AllToAll. The feature supported in HetuMoE is compared with other MoE systems in Figure~\ref{fig:sys_compare}.

\subsection{Gating Strategy in HetuMoE}
\paragraph{Top1/Top2/Topk} Shazeer et al., 2017~\cite{shazeer2017outrageously} proposed to utilize Topk activated MoE layer in LSTM increase the model capacity up to $1000\times$ with only minor losses in efficiency. The Topk gate is formulated in Equation~\ref{equ:topkgate}, where $x$ is the $N$ input tokens, $W$ is the gate's weight, E experts $e_{i}(i \in {1..E})$ and $y$ is the output tokens.
GShard~\cite{GShard} and Switch~\cite{fedus2021switch} propose to simplify the gate to Top2 and Top1, and utilize a capacity factor $C$ to force the max received tokens by each expert.
As the $K$ increases, the MoE layer tends to perform better while leading to more computation. It's a trade-off on $K$ considering computation efficiency and model performance.

\begin{equation}
\begin{aligned}
    g = softmax(TopK(x \cdot W, K) ) \\
    y = \sum_{i=1}^{E}g(x)_{i} \cdot e_{i}(x)
\end{aligned}
\label{equ:topkgate}
\end{equation}

\begin{figure}[t]
    \centering
        \subfigure[Different Expert Number]{
            \scalebox{0.52}{\includegraphics[width=0.75\columnwidth]{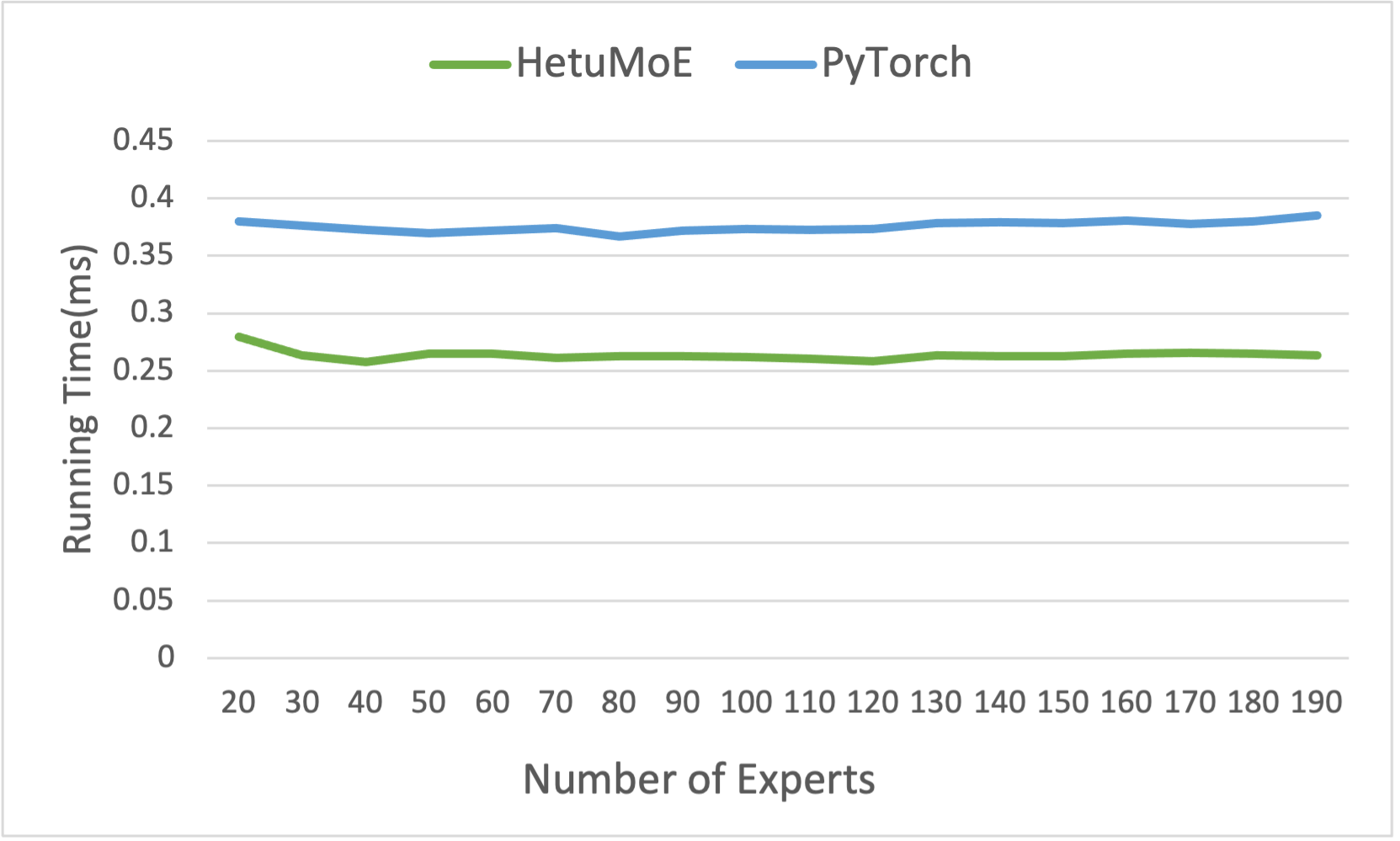}}
        }
        \subfigure[Different Token Number]{
            \scalebox{0.52}{\includegraphics[width=0.75\columnwidth]{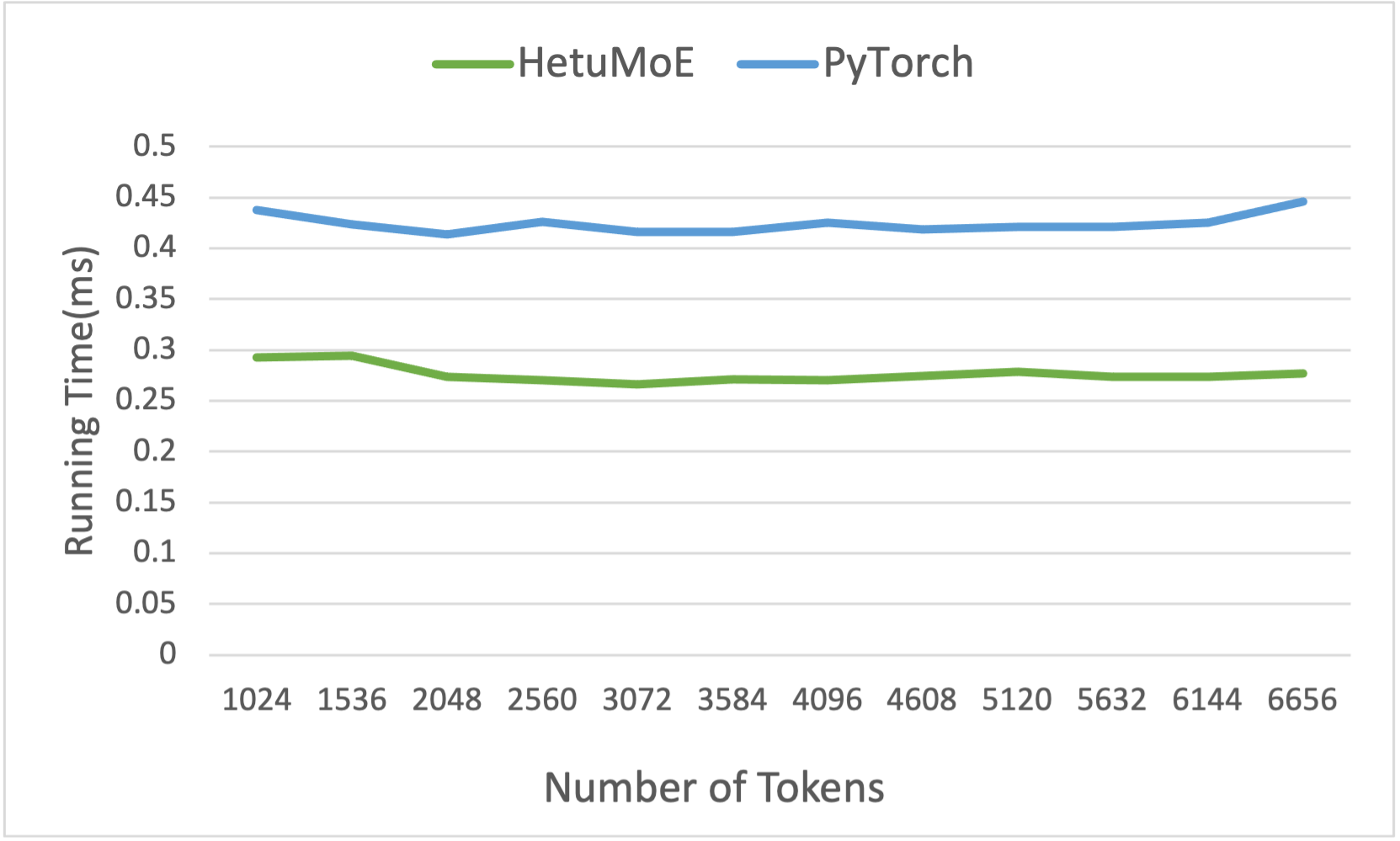}}
        }
    \caption{Topk kernel performance comparison with PyTorch} 
\label{fig:topk_eval}
\end{figure}

\paragraph{kTop1}
Inspired by the observation that $K$ and expert capacity $C$ in Topk routing can significantly make a difference in model performance, M6-T~\cite{M6T} propose the KTop1 gating strategy, where experts are divided into $k$ prototypes and each token is assigned to the highest scored expert in each prototype. Finally, the outputs of different prototypes are summed together as for the same input token.
\paragraph{Hierarchical Topk(H Topk)}
As increasing the number of activated experts can boost the model performance with a higher sparse ratio, SAM~\cite{SAM} (Switch and Mixture) proposed an efficient hierarchical routing mechanism that divides the experts into different groups according to their devices and activates multiple experts in the same group to avoid the remote communication cost between devices. Specifically, the \textit{Switch Router} first selects one group and then the \textit{Mixture Router} selects multiple experts in the same group for each token.


\paragraph{BASE Layer} 
BASELayer~\cite{BaseLayer} formulates token-to-expert allocation as a linear assignment problem to improve efficiency, where balanced loads are guaranteed among each expert. Because no new parameters or auxiliary expert-balanced losses are introduced, the training process is simplified. The problem is formulated in Equation~\ref{equ:base}, where $N$ tokens with representations $x_{i}$ and $E$ experts with embeddings $w_e$, we assign each token to an expert via the assignment index $a_{i} \in 0..E$:
\begin{equation}
\begin{aligned}
max \sum_{i=1}^{N}x_{i} \cdot w_{a_{t}} \\
s.t.\ \forall{e}\sum_{i=1}^{N}1_{a_{i}=e} = \frac{N}{E}
\label{equ:base}
\end{aligned}
\end{equation}


\paragraph{Hash Layer} Hash Layer~\cite{HashLayer} utilizes various hash functions to map tokens to experts, such as Random Hashes, Balanced assignment, and Clustered Hashes. Specifically, 
in Hash Layer, tokens are placed in the corresponding buckets according to the used hash function and each bucket refers to an expert.

\begin{figure}[h]
    \centering
        \subfigure[Transformation]{
            \scalebox{0.32}{\includegraphics[width=0.75\columnwidth]{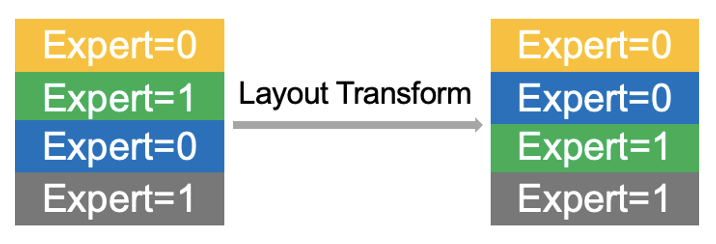}}
        }
        \subfigure[Reverse transformation]{
            \scalebox{0.32}{\includegraphics[width=0.75\columnwidth]{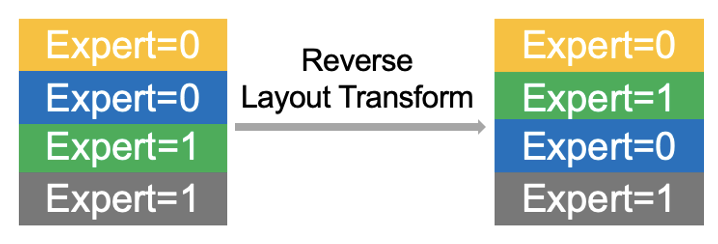}}
        }\\
        \subfigure[Performance comparison with Tutel]{\scalebox{0.48}{\includegraphics[width=\columnwidth]{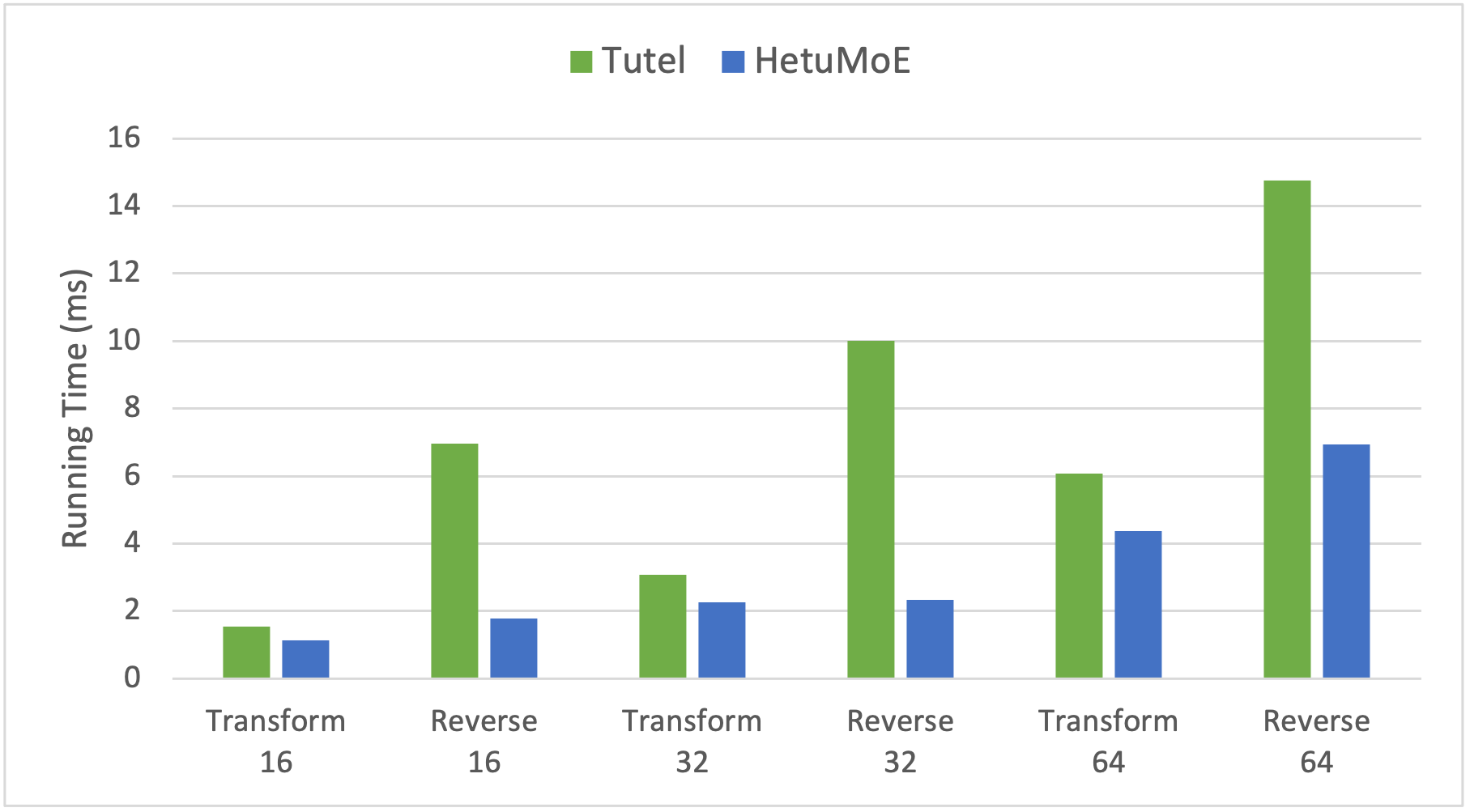}}
        }
    \caption{Data layout transformation illustration and performance comparison} 
\end{figure}

\paragraph{Dense-To-Sparse} Dense-To-Sparse Gate~\cite{dense_to_sparse} considers that current approaches of jointly training experts and the sparse gate introduce a negative impact on model accuracy, diminishing the efficiency of expensive large-scale model training~\cite{fedus2021switch}. It proposes to begin as a dense gate that routes tokens to all experts, then gradually and adaptively become sparser while routing to fewer experts. Specifically, it utilizes the Gumbel Softmax and decreases the temperature during training.

\subsection{HetuMoE Optimization}
As shown in Algorithm~\ref{alg:moe_training}, \texttt{Gate}, \texttt{Layout Transform} and \texttt{AllToAll} are three key components related to MoE models' training, where the module of expert networks also exists in common models and thus is not our target in MoE-specific optimization. The optimizations in HetuMoE about these three modules are detailed as follows.
\paragraph{Gate Optimization} Various gating strategies are supported in HetuMoE and we mainly optimize the $topk$ operator as it is widely adopted in mainstream models~\cite{fedus2021switch,GShard,vmoe}. The operator's input is a 2-D tensor which has shape $(num\_tokens, num\_experts)$ and executes as finding the top-k values and the corresponding index for each row. 
The Topk operator implemented in PyTorch or Tensorflow for arbitrary K. We conduct algorithmic optimizations for useful $k$ in MoE, such as 1 and 2. We vary $num\_experts$ and $num\_tokens$ to conduct comparison with PyTorch and the experimental results are shown in Figure~\ref{fig:topk_eval}. We outperform PyTorch Topk by $25\%$ speed improvement averagely.


\paragraph{Layout Transform Optimization}
Data layout transformation is another important step in the MoE training process. After the gating network decides the token-to-expert mapping, tokens assigned to the same expert need to be put in physically continuous memory locations.
We undertake kernel level optimization to accelerate this process, and we achieve more than 26\% improvement compared with state-of-art implementation.

\begin{figure}[t]
    \centering
    \includegraphics[width=0.75\columnwidth]{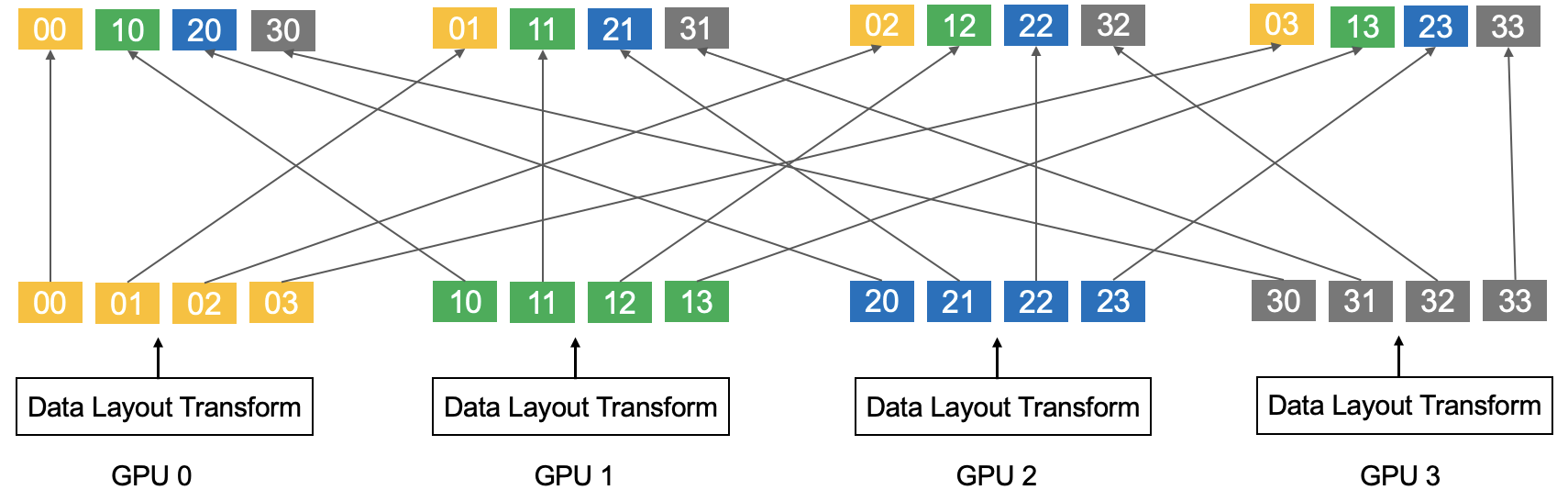}
    \label{fig:alltoall}
    \caption{Illustration of NCCL AllToAll.}
\end{figure}

\paragraph{All-To-All Optimization} In AllToAll operation, each GPU sends its data to all GPUs (one-for-all) and receives data sent by all GPUs (all-for-one), where each data will be divided equally into n parts, illustrated in Figure~\ref{fig:alltoall}.
Current AllToAll operations implemented in NCCL and MPI may suffer from low utilization of network bandwidth because of the small message size. Specifically, if we have N nodes, each with G GPUs, and each GPU has data size B, then the size of data transferred between 2 GPUs is $\frac{B}{GN}$. A common setting is $N=8$, $G=8$, and $B=16$MB. 

\begin{figure}[t]
\centering 
\includegraphics[width=0.8\columnwidth]{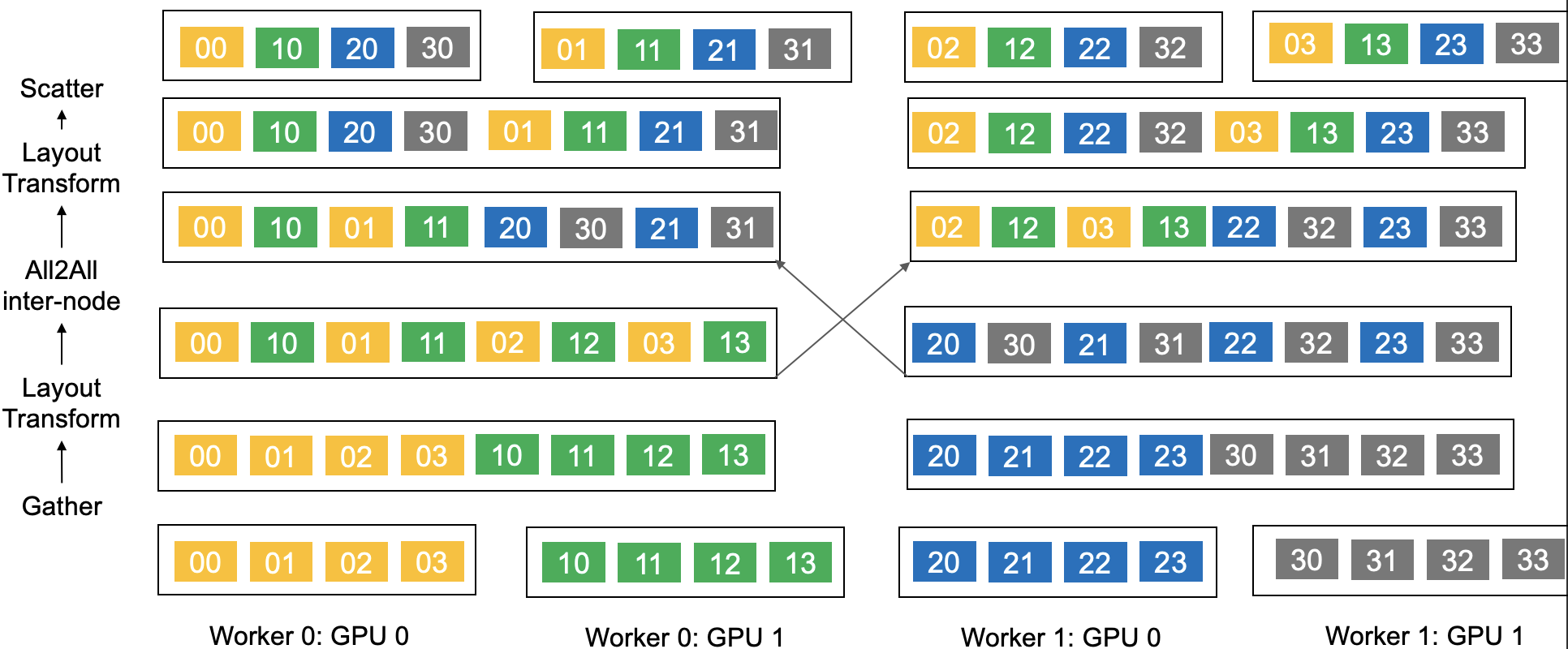}
\caption{Illustration of hierarchical AllToAll}
\label{fig:halltoall}
\end{figure}

\begin{figure}[t]
    \centering
        \subfigure[4 nodes]{
            \scalebox{0.53}{\includegraphics[width=0.75\columnwidth]{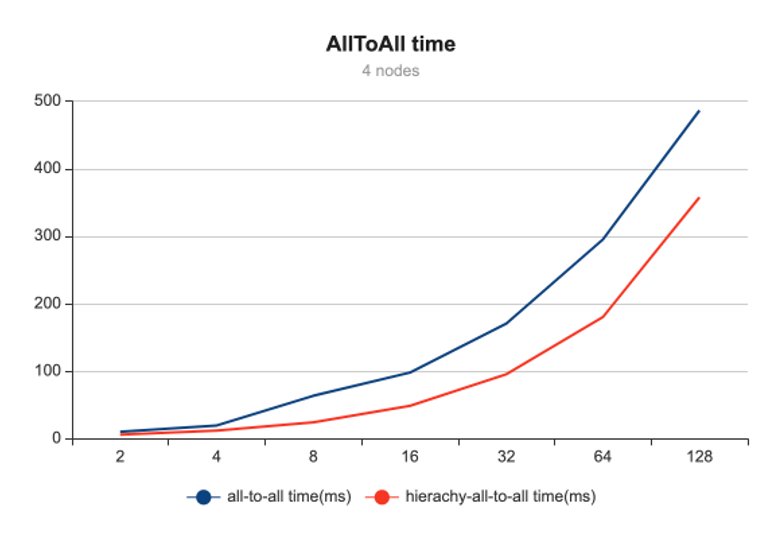}}
        }
        \subfigure[8 nodes]{
            \scalebox{0.53}{\includegraphics[width=0.75\columnwidth]{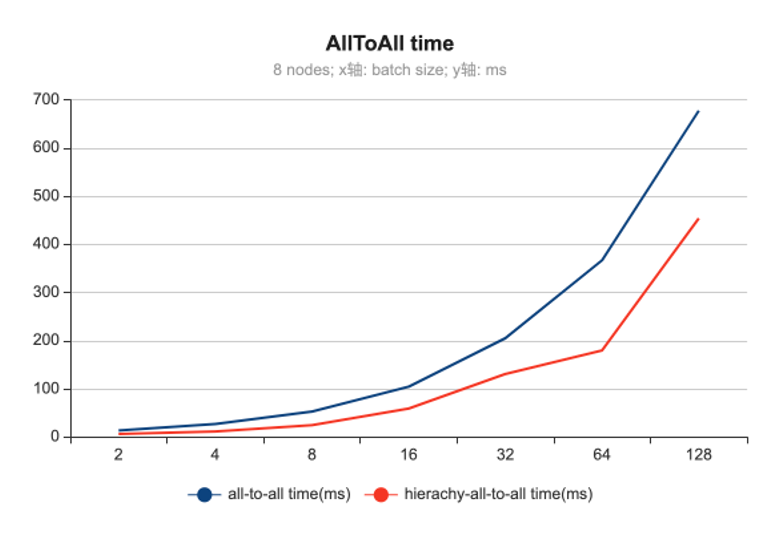}}
        }
    \caption{Hierarchical AllToAll performance} 
\end{figure}

In HetuMoE, we utilize \texttt{Hierarchical AllToAll}, which combines hierarchical networks (intra-node and inter-node) and aggregates messages, to optimize the communication between multi-nodes equipped with one NIC.
Jesper et al., 2014~\cite{ha2a} also utilize hierarchy to optimize alltoall communication for MPI.
Illustrated as Figure~\ref{fig:halltoall}, it first gathers the data of all GPUs inside one node into one GPU. Then, a data layout transformation is undertaken to place the token assigned to the same node in physically continuous memory. Afterwards, it launches All2All communication between nodes. After AllToAll is done, it conducts the corresponding data layout transformation and scatter operation to put each token to its corresponding expert. In this way, the size of data transferred between nodes becomes $\frac{BG}{N}$, which is $G^2$ times larger than before. 
Meanwhile, this two-level decoupled AllToAll also fully utilizes the intra-node (NvLink or PCIe) and inter-node bandwidth (Infiniband or Ethernet).
Experiments show that \texttt{Hierarchical AllToAll} achieves $1.66\times$ speedup in $4\times8$ GPUs setting and $2\times$ speedup in $8\times8$ GPUs setting.

\begin{figure}[t]
    \centering
        \subfigure[Switch Gate]{
            \scalebox{0.53}{\includegraphics[width=0.75\columnwidth]{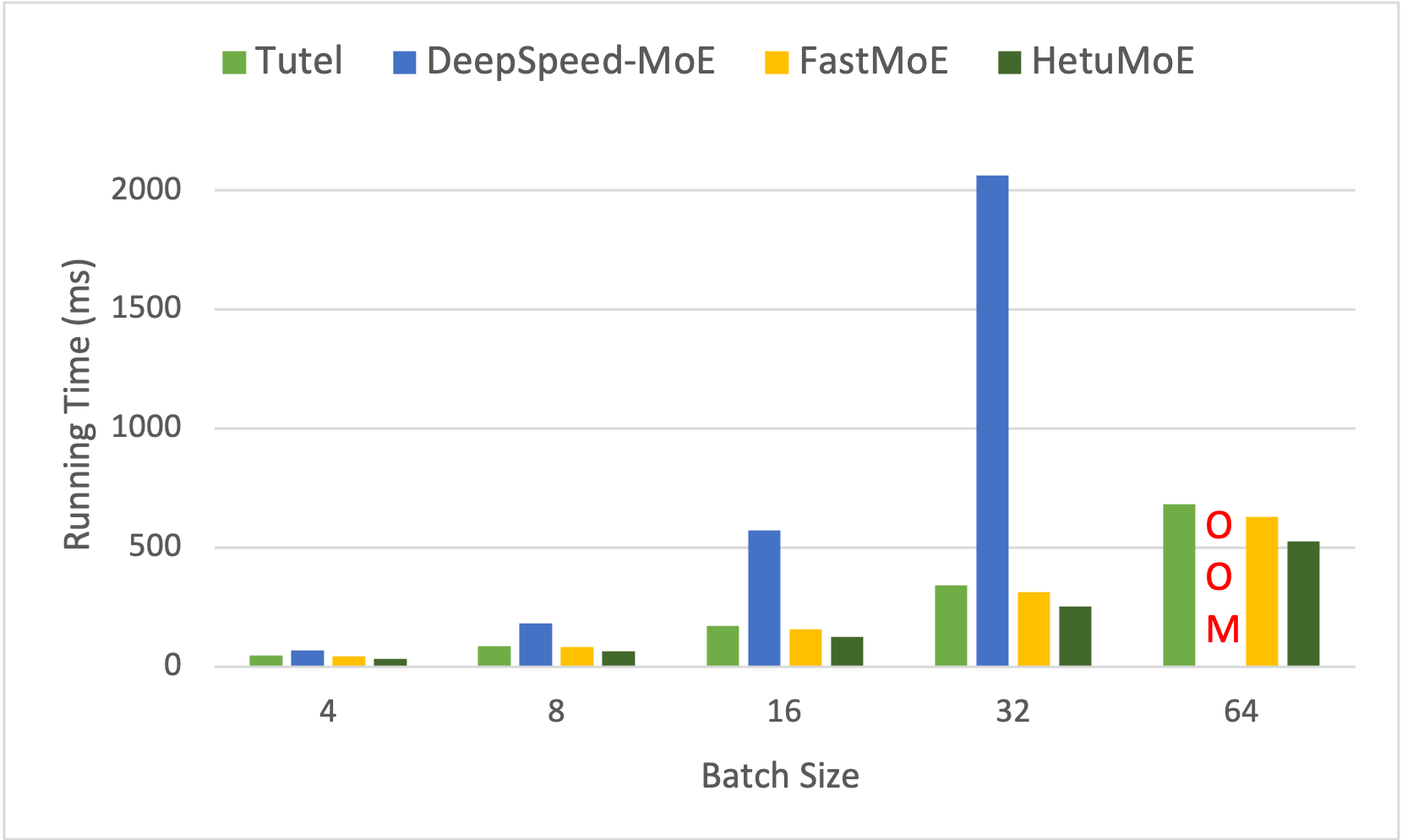}}
        }
        \subfigure[GShard Gate]{
            \scalebox{0.53}{\includegraphics[width=0.75\columnwidth]{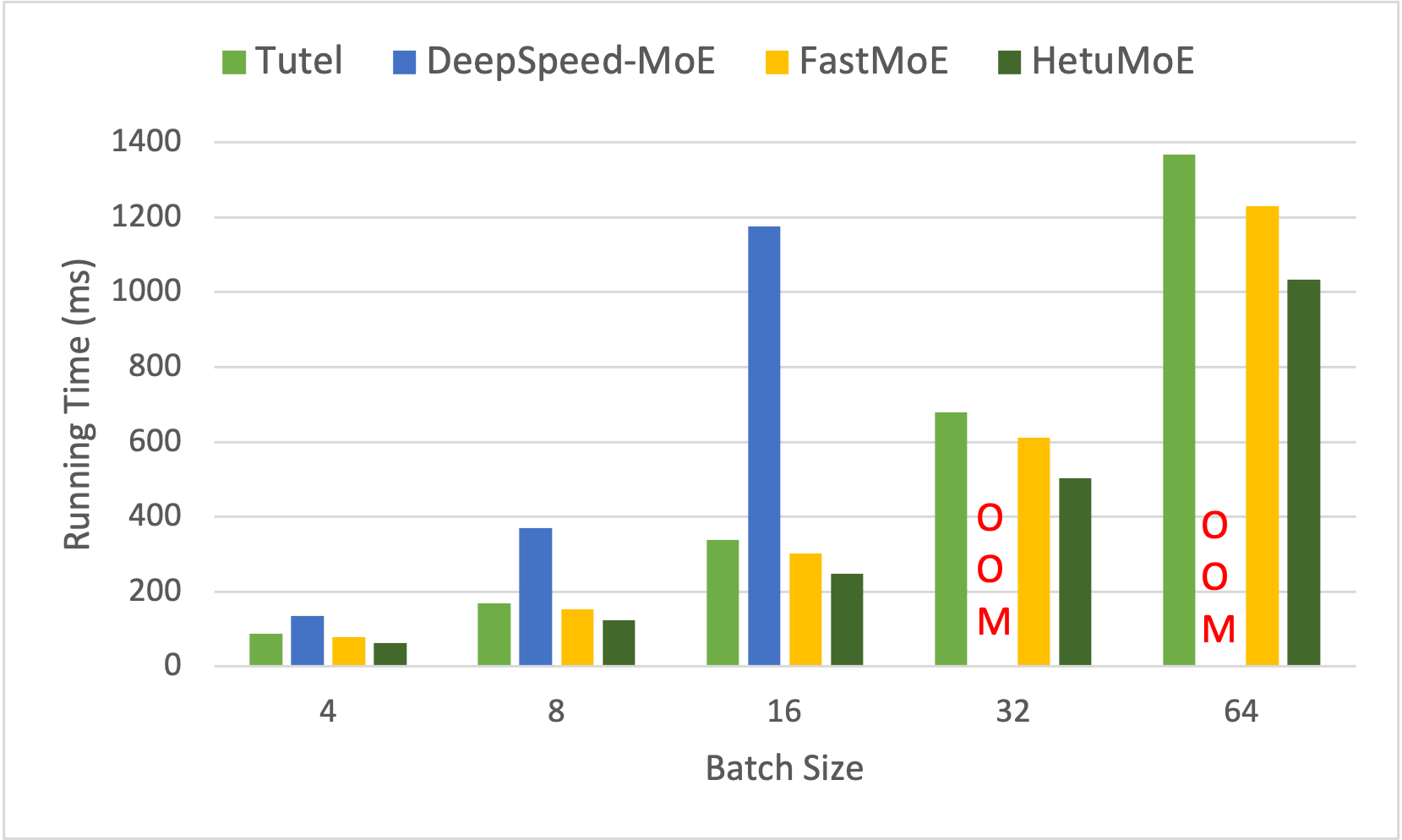}}
        }
    \caption{Overall performance comparison with DeepSpeed-MoE, FastMoE, and Tutel.} 
    \label{fig:overall}
\end{figure}

\paragraph{Overall Performance}
Experiments are conducted on GPU Clusters, where each node is equipped with 8 TITAN RTX GPUs and these eight GPUs are connected through PCIe.
We compared HetuMoE with existing MoE systems, including Tutel, FastMoE and DeepSpeed-MoE. 
Our test model is a 16-experts MoE layer, where each expert represents a FeedForward Network and the hidden size is 2048, The sequence length and embedding dimension of input data is 1024 and 2048 respectively. We vary a batch size to compare the performance of each system.

As shown in Figure~\ref{fig:overall}, HetuMoE achieves the state-of-art training performance in both Switch gate and GShard gate.
Specifically, compared with FastMoE, we achieve $18\%$ speed-up in Switch Gate and $15\%$ speed-up in GShard gate. Meanwhile, HetuMoE outperforms DeepSpeed-MoE up to $8.1\times$ under the switch gate with a batch size of 32. 


\section{Conclusion}
Existing MoE training systems lack various gating strategy support and perform badly in distributed training.
In this paper, we propose HetuMoE, a high-performance distributed MoE training system built on Hetu, which supports a variety of mainstream gating strategies and achieves state-of-the-art training speed compared to existing baseline systems.
HetuMoE adopts several customized gating kernels' implementation and utilizes the \texttt{hierarchical All-To-All} to optimize distributed communication by combining hierarchical network (intra-node and inter-node) and aggregating small messages.

\bibliographystyle{abbrv}
\bibliography{reference}

\end{document}